\documentclass[twocolumn,aps,amsmath,amssymb,prb]{revtex4}
\usepackage{graphicx}
\usepackage{epsfig}
\bibstyle{apsrev.bib}

\begin{document}
\input epsf.sty

\title{Entanglement entropy with localized and extended interface defects}
\author{Ferenc Igl\'oi}
\affiliation{ Research Institute for Solid
State Physics and Optics, H-1525 Budapest, P.O.Box 49, Hungary}
\affiliation{
Institute of Theoretical Physics,
Szeged University, H-6720 Szeged, Hungary}
\author{Zsolt Szatm\'ari}
\affiliation{
Institute of Theoretical Physics,
Szeged University, H-6720 Szeged, Hungary}
\author{Yu-Cheng Lin}
\affiliation{
Theoretische Physik, Universit\"at des Saarlandes, D-66041
Saarbr\"ucken, Germany }

\date{\today}

\begin{abstract}
The quantum Ising chain of length, $L$, which is separated into
two parts by localized or extended defects is considered at the critical
point where scaling of the interface magnetization is non-universal. We measure
the entanglement entropy between the two halves of the system in equilibrium,
as well as after a quench, when the interaction at the interface is changed for time $t>0$.
For the localized defect the increase of the entropy with $\log L$ or with $\log t$ involves the
same effective central charge, which is a continuous function of the strength of the defect.
On the contrary for the extended defect the equilibrium entropy is saturated, but the non-equilibrium
entropy has a logarithmic time-dependence the prefactor of which depends on the strength of the defect.
\end{abstract}

\maketitle

\newcommand{\bc}{\begin{center}}
\newcommand{\ec}{\end{center}}
\newcommand{\be}{\begin{equation}}
\newcommand{\ee}{\end{equation}}
\newcommand{\beqn}{\begin{eqnarray}}
\newcommand{\eeqn}{\end{eqnarray}}


\vskip 2cm

\section{Introduction}

Entanglement, quantum nonlocality and quantum correlations has became the subject
of intensive research recently in different fields
of physics\cite{Amicoetal08}: quantum information theory, condensed matter physics, quantum
field theory, etc. For a quantum system which is divided into two
parts, $\cal A$ and $\cal B$, all
information about entanglement is encoded in the reduced density matrix: $\rho_{\cal A}=Tr_{\cal B} |\Psi\rangle \langle \Psi |$, where $|\Psi \rangle$ is a pure state of the complete system. The entanglement
between $\cal A$ and $\cal B$ is conveniently measured by the von Neumann
entropy $S_{A}=-\textrm{Tr}_{A} (\rho_{\cal A} \log \rho_{\cal A})$, which has been intensively
studied in many-body systems, in particular in one dimension (1d). For a
critical 1d system (with periodic boundary conditions)
the entropy is found to grow logarithmically with the length, $L$:
\be
S_{A}= \frac{c}{3} \log L + c_1\;,
\label{S_l}
\ee
where $L$ is the size of $\cal A$ or the size of the complete system, provided it is
divided into two equal parts\cite{Holzhey,Latorre03,CC04}. For conformally invariant systems the
parameter in the prefactor,
$c$, is universal and given by the central charge of the conformal algebra. In the vicinity
of the critical point where the correlation length is $\xi \ll L$, the entropy is saturated
and given by:
\be
S_{A}\simeq \frac{c}{3} \log \xi \;.
\label{S_xi}
\ee
One is also interested in the time evolution of the entropy\cite{CC05} after changing the form of
the interaction (quantum quench) at time $t=0$. In the case of a local quench\cite{EP07} the interaction
parameters are modified in a restricted region. For example measuring the entropy between
$\cal A$ and $\cal B$ which are
disconnected for $t<0$ but are joined to a closed chain with homogeneous couplings for $t>0$
at the critical point we observe a logarithmic increase in time, $t \ll L$, as\cite{CC07,EKPP08}
\be
S_{A}= 2\frac{c}{3} \log t +{\rm cst.}\;.
\label{S_t}
\ee
If the complete system is open, i.e. there is one boundary point between $\cal A$ and $\cal B$
the prefactors in Eqs.(\ref{S_l}-\ref{S_t}) are divided by a factor $2$.

Inhomogeneous interactions could modify the entanglement properties of quantum spin chains.
It has been shown that for
random\cite{RefaelMoore04,Laflo05,DeChiara,Santa06,RefaelMoore07,BonesteelYang07,ilrm07,IgloiLin08} and
aperiodic\cite{ijz07}
couplings the prefactor in Eq.(\ref{S_l}) is changed and involves
the so called effective central charge, $c_{eff}$. On the other hand if the couplings
vary linearly with the position an interface with a certain width is introduced, and
in the expression of the entropy in Eq.(\ref{S_xi}) $\xi$ is replaced by this length\cite{eip09}. 

If the inhomogeneities are centered at a few points ("defects") they are not expected to modify the
scaling form of the entropy, unless the defects are located at the
interface. Indeed interface defects can modify the scaling form of the
wavefunction in the vicinity of the junction\cite{ipt}, which in turn alter the entanglement entropy.
The effect of a local interface defect, $\Delta$, which measures the coupling
between $\cal A$ and $\cal B$ has been investigated for $XXZ$ and $XX$ quantum spin
chains\cite{ZhaoPeWa06,Peschel05,Levine04}.
For the antiferromagnetic $XXZ$ chain the defect is a (marginally) relevant perturbation\cite{EggertAffleck92},
the defect renormalizes to a cut and the effective central charge approaches zero\cite{ZhaoPeWa06}. On the
contrary for the ferromagnetic $XXZ$ chain the defect is a (marginally) irrelevant perturbation,
the defect renormalizes to the homogeneous coupling and the effective central charge approaches one.
Finally, in the $XX$ chain the defect is a marginal perturbation and the effective central
charge in Eq.(\ref{S_l}) is found\cite{Peschel05} to depend on the strength of the defect, $\Delta$.

In the present paper we study the effect of interface defects on the entanglement
properties of critical quantum spin chains. Our approach differs from the previous ones in
several respects. The system we consider is the quantum Ising chain and we study the problem
with a localized as well as with an extended defect. The latter is realized by a smooth inhomogeneity
in the couplings varying as $\simeq A/x$, $x$ being the distance from the
interface\cite{HilhorstLeeuwen81,ibt90}.
Both perturbations
are known to be marginal as far as the scaling behavior of the interface magnetization at the critical
point is considered. We study the entropy both in equilibrium, as well as after a quench,
when the interface couplings are modified for $t>0$.
The main goal of our investigations is to study possible relations i) between local critical scaling
and the scaling of the entropy, and ii) between scaling form of equilibrium and non-equilibrium entropies
like in Eqs. (\ref{S_l}) and (\ref{S_t}).

The structure of the paper is the following. The model, the type of defects, the interface critical
behavior as well as the way of calculation of the entropy is described in Sec.\ref{sec:model}. The
localized and the extended defect problems are studied in Sec.\ref{sec:localised} and
\ref{sec:extended},
respectively. Our results are discussed in the final Section. Some technical details of the calculations
are put in Appendices.

\section{Models and method}
\label{sec:model}
We consider the quantum Ising chain defined by the Hamiltonian:
\be
H_0 =
-\sum_{i=1}^L \sigma_i^x \sigma_{i+1}^x- h \sum_{i=1}^L \sigma_i^z
\label{eq:H0}
\ee
in terms of the Pauli-matrices, $\sigma_i^{x,z}$, at site $i$ and with periodic
boundary conditions, $\sigma_{L+1}^x=\sigma_{1}^x$. The quantum critical
point of the system is given by\cite{pfeuty} $h=h_c=1$, where the bulk correlation function has
a power-law decay for large $L$:
\be
G(L)=\langle 0 | \sigma_{L/4}^x \sigma_{3L/4}^x | 0 \rangle\ \sim L^{-\eta}\;,
\label{GL}
\ee
with $\eta=\eta_0=1/4$.

\subsection{Localized defect}

A localized defect is defined by the perturbation:
\be
V_{loc} =
(1-\Delta) (\sigma_{L/2}^x \sigma_{L/2+1}^x+\sigma_{L}^x \sigma_{L+1}^x)
\label{eq:Hloc}
\ee
so that the complete Hamiltonian is given by $H_0+V_{loc}$. This perturbation
does not modify the decay of the bulk correlations in Eq.(\ref{GL}), however the interface
or defect correlations:
\be
G_d(L)=\langle 0 | \sigma_{1}^x \sigma_{L/2}^x | 0 \rangle\ \sim L^{-\eta_d}\;,
\label{GdL}
\ee
involve a new exponent\cite{bariev,McCoyPerk}:
\be
\eta_d=\eta_{loc}(\Delta)=\frac{4}{\pi^2} \arctan^2(1/\Delta)
\label{eta_d}
\ee
which is a continuous function of the strength of the defect, $\Delta$.
Note that in the special cases,
$\Delta=0$ and $\Delta=1$ we recover the known results for surface and bulk correlations, respectively.

More generally, one can consider the defect at position $L$ of different strength, say $\Delta'$. For
$\Delta'=1$ and for $\Delta'=0$ there is one defect in the closed or open chain, respectively. The
defect exponent is then modified to $(\eta_d+1/4)/2$ and $(\eta_d+1)/2$, respectively\cite{ipt}.

\subsection{Extended defect}

The extended defect is defined by a smooth inhomogeneity\cite{ibt90}:
\be
V_{ext} =
-\sum_{i=1}^L \lambda_i \sigma_i^x \sigma_{i+1}^x,\quad 
\lambda_i=\frac{A/2}{\frac{L}{2\pi} \left|\sin \left[ \frac{2\pi(i-\delta)}{L} \right]\right|}
\label{eq:Hext}
\ee
and this perturbation is put symmetrically in the two parts of the lattice.
In Eq.(\ref{eq:Hext}) we have used a shift, $\delta=O(1)$,
in order to avoid singularities.

The local critical behavior of this system is different for $A<0$ and for $A>0$.
For weakened local couplings, $A<0$, correlations between two defect spins, $G_d(L)$,
has an algebraic decay with an $A$ dependent defect exponent: $\eta_{ext}(A)=1-A$. At
the same time correlations between two bulk spins, $G(L)$, involves an exponent $\eta$
which also depends on $A$: $\eta=1/4-A$. This unexpected variation of the
exponent with $A$ is due to
the fact that the interface coupling at $i=0$ (as well as at $i=L/2$) renormalizes
to zero as $J_0(L) \sim L^{A}$. This fact explains also the observation that the decay exponent of
the end-to-end correlations for two decoupled chains is just $\eta_{ext}(A)$, at least for $A<0$.

For enhanced local couplings, $A>0$, the defect-defect correlations, $G_d(L)$, approach a
finite limiting value, $m_d^2$, so that the interface stays ordered at the bulk critical point. The
connected correlation function, $G_d(L)-m_d^2$, decays to zero algebraically with an
exponent $\eta'_{ext}(A)=2A$. In this case correlations between bulk spins, $G(L)$, involves
the pure exponent, $\eta_0=1/4$.

\subsection{Calculation of the entropy}

Calculation of the entropy of the quantum Ising chain in the equilibrium case is described
in detail in several papers\cite{Peschel03,Latorre03,IgloiLin08}.
The nonequilibrium entropy in the homogeneous case and
in the thermodynamic limit is calculated in\cite{CC05}, whereas for inhomogeneous couplings
the method is described in the Appendix \ref{sec:quadr}. Here we briefly recapitulate the
main steps of the calculation and describe the technical steps needed in the numerical calculation.

The first step is to transform the Hamiltonian of the quantum Ising chain in terms of free
fermions\cite{lsm}. Numerically, this step
necessitates the diagonalization of an $L \times L$ symmetric matrix. In the second
step we calculate the reduced density matrix, which can be reconstructed from the correlation matrix
in the free fermionic basis\cite{Peschel03}. The entanglement entropy is calculated then from the eigenvalues of the
reduced correlation matrix., If ${\cal A}$ has $\ell$ sites (in
our case we use $\ell=L/2$) this second step requires the diagonalization of a $\ell \times \ell$
symmetric matrix, if one works with Dirac fermions\cite{IgloiLin08} or a $2\ell \times 2\ell$
skew-symmetric matrix
if the calculation is performed with Majorana fermions\cite{Latorre03}.
At this step we have calculated the so called single-copy entanglement\cite{PeschelZhao}, too,
which is defined through the largest eigenvalue of the density matrix, $w_1$, as $S_1=-\log w_1$.
$S_1$ is obtained also from the eigenvalues of the correlation matrix. For homogeneous
chains $S_1(L)$ is known to have a logarithmic size-dependence:
\be
S_1= \frac{\kappa}{3} \log L + {\rm cst}\;,
\label{S_1}
\ee
with a prefactor: $\kappa=c/2$.

In the third step of the calculation we consider the time evolution of the entropy after a quench.
For this we should calculate the time evolution of the correlation matrix\cite{IgloiRieger00},
which can be done in the Majorana fermion basis. Each matrix-elements is obtained through $L^2$
operations, which will result in the increase of the computational time accordingly.

If the entropy has a logarithmic dependence, either as a function of the size (see Eq.(\ref{S_l}))
or the time (see Eq.(\ref{S_t})), then we have calculated the prefactors through two-point fit and
in this way effective central charges are obtained as:
\be
c_{eff}(L)=3[{\cal S}(2L)-{\cal S}(L)]/\log 2\;.
\label{two_fit}
\ee
Similarly in the nonequilibrium case we calculated the prefactors as:
\be
p(L)=[{\cal S}(2L,t=L/2)-{\cal S}(L,t=L/4)]/\log 2\;.
\label{two_fit_t}
\ee
From this series of results an estimate of $c_{eff}$ or $p$
is obtained by sequence extrapolation methods, such as by the Bulirsch-Stoer algorithm\cite{BulirschStoer}.
In the numerical calculation we used finite systems up to $L=1024$ for the equilibrium entropy
and up to $L=512$ for the non-equilibrium entropy.

\section{Chain with localized defects}
\label{sec:localised}
\begin{figure}[h]
\begin{center}
\includegraphics[width=3.3in,angle=0]{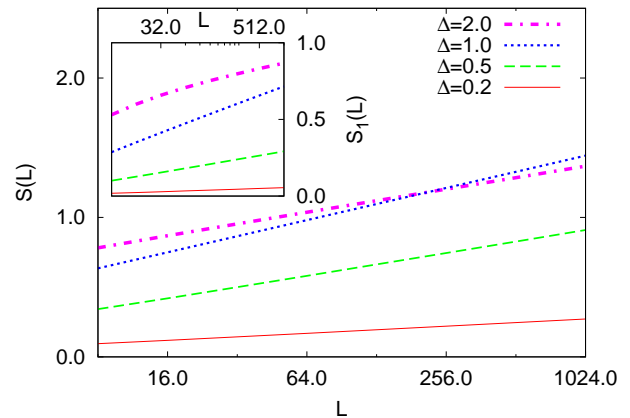}
\end{center}
\caption{
\label{fig:Sloc} (Color online)
Entanglement entropy of the quantum Ising
chain with different strength of the defect
as a function of the logarithm of the length. Inset: the same for the single-copy entanglement.}
\end{figure}

Having two symmetrically placed defects of strength $\Delta$ in the quantum Ising chain,
as given by the Hamiltonian $H_0+V_{ext}$ in Eqs.(\ref{eq:H0}) and Eq.(\ref{eq:Hloc})
we have calculated the entanglement entropy between two halves of the system for
different lengths, $L$. The results are shown in Fig.\ref{fig:Sloc} as a function of
$\log L$. 

%
\begin{figure}[h]
\begin{center}
\includegraphics[width=3.3in,angle=0]{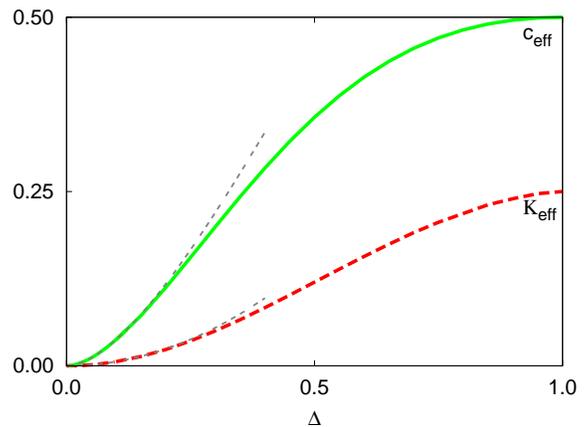}
\end{center}
\caption{
\label{fig:loc} (Color online)
Effective central charge, $c_{eff}(\Delta)$ (full line - green), and the parameter, $\kappa_{eff}(\Delta)$
(broken line - red),
of the single-copy entanglement of the quantum Ising chain as a function of the strength of
the localized defect, $\Delta$. For
small $\Delta$ the leading behaviors obtained by perturbational calculation are indicated by dotted lines.}
\end{figure}

%
For large $L$ the curves approach straight lines with different $\Delta$-dependent
slopes. We have calculated effective central charges by two point fits, see Eq.(\ref{two_fit}),
and their extrapolated values are put in Fig. \ref{fig:loc}. 
%
%
These are continuous function of the
strength of the defect and vary from $0$ to $1/2$ as $\Delta$ tuned from $0$ to $1$. The
numerical data are consistent with the relation, $c_{eff}(\Delta)=c_{eff}(1/\Delta)$, which
symmetry holds also for the local magnetization exponent in Eq.(\ref{eta_d}). For small $\Delta$
we have calculated ${\cal S}(L)$ perturbatively, the calculation is presented in Appendix \ref{sec:pert}.
We have obtained that in leading order of $\Delta^2$, ${\cal S}(L)$ has a logarithmic $L$-dependence
and the effective central charge is given by:
\be
c_{eff}(\Delta)=6 \Delta^2 \left(\frac{1}{\pi^2}(1-\ln \Delta^2) + b\right)+O(\Delta^4)\;.
\label{c_eff_Delta}
\ee
Here $b=0.062180(2)$ is a numerically calculated constant. The perturbative result in Eq.(\ref{c_eff_Delta})
is also shown in Fig.\ref{fig:loc} together with the numerical data. It is interesting to note
that in Eq.(\ref{c_eff_Delta}) there is a logarithmic correction to the leading $\Delta^2$ behavior.

We have also calculated the single-copy entanglement, ${\cal S}_1$, which is found to be in the form of
Eq.(\ref{S_1}), however with $\Delta$ dependent prefactors, $\kappa_{eff}(\Delta)$, the extrapolated
values of which are plotted in Fig.\ref{fig:loc}, too. In the range $0<\Delta<1$,
$\kappa_{eff}(\Delta)$ is seen to vary between $0$ and $1/4$ and in the small $\Delta$ limit
we have (see Appendix \ref{sec:pert}):
\be
\kappa_{eff}(\Delta)= \frac{6 \Delta^2}{ \pi^2}+O(\Delta^4)\;.
\label{kappa_eff}
\ee
The ratio $\kappa_{eff}(\Delta)/c_{eff}(\Delta)$ varies between $0$ and $1/2$, thus the
conformal result, $\kappa/c=1/2$ is valid only in the homogeneous system.

Our results can be generalized if there is one defect in the system. For the closed chain
(with $\Delta'=1$) the effective
central charge is changed to $(c_{eff}(\Delta)+1/2)/2$, whereas for an open chain (with $\Delta'=0$)
the prefactor of the logarithm is just the half as for two defects.

\subsection{Time evolution after the quench}

\begin{figure}[h]
\begin{center}
\includegraphics[width=3.3in,angle=0]{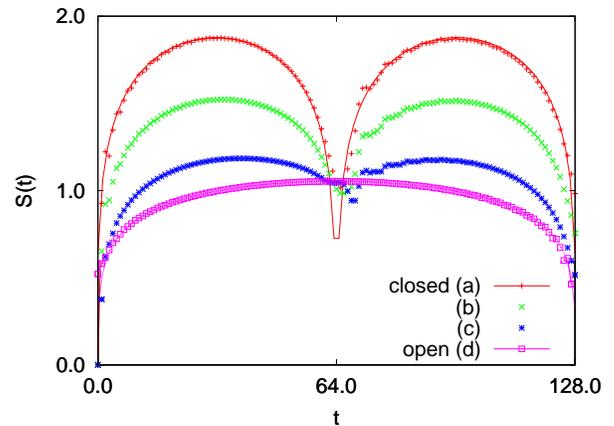}
\end{center}
\caption{
\label{fig:S_t_loc} (Color online)
Time evolution of the entropy after a quench from two disconnected chains. The final
system is closed and homogeneous (a) or contains one (b) or two (c) defects of strength $\Delta=1/2$.
The quench to a homogeneous open chain is given by (d). The analytical formulae for the
homogeneous closed Eq.(\ref{S_t_conf}) and open Eq.(\ref{S_t_conf1}) chains are also shown by full lines.}
\end{figure}

Here we calculate the time evolution of the entropy by starting with two disconnected
parts, i.e. with $\Delta=0$ (and $\Delta'=0$) for $t<0$, and connecting them by one or two defects
with $\Delta>0$ for $t>0$. In a finite chain of length $L$ the entropy has a periodic time-dependence,
the period of which is $L/2$, if the final chain is closed ($\Delta'=\Delta$ or $\Delta'=1.$)
and the period is $L$, if the final chain is open ($\Delta'=0.$). This is illustrated in Fig.\ref{fig:S_t_loc}
for a chain with $L=128$ sites and with different type of defects.
%
%

If we perform the quench to
the homogeneous chain, the numerical data are very well fitted by the formula:
\be
{\cal S}_L^{cl}(t)=2 \frac{c}{3} \log \left| \frac{L}{2 \pi} \sin \frac{2 \pi t}{L} \right|+{\rm cst}\;,
\label{S_t_conf}
\ee
for a closed chain and
\be
{\cal S}_L^{op}(t)=\frac{c}{3} \log \left| \frac{L}{ \pi} \sin \frac{ \pi t}{L} \right|+{\rm cst}\;,
\label{S_t_conf1}
\ee
for an open chain with $c=1/2$, which are also shown in Fig.\ref{fig:S_t_loc}.

In the general case, when the closed chain for $t>0$ contain one or two
defects of strengths $0<\Delta < 1$, the numerical
results in Fig.\ref{fig:S_t_loc} are compatible with the scaling form (for closed chains):
\be
{\cal S}_L^{cl}(t)=2 \frac{c_{eff}}{3} \log\left[ L f^{cl}\left( \frac{ t}{L}\right)\right] +{\rm cst}\;,
\label{S_t_scal}
\ee
where the scaling function, $f^{cl}(y)>0$, is periodic with period $1/2$ and for small argument
it behaves as $f^{cl}(y) \sim y$. For an open chain with one defect the prefactor in Eq.(\ref{S_t_scal})
is changed to its half and the scaling function, $f^{op}(y)>0$, is periodic with period $1$.
Then for $t \ll L$ we expect the asymptotic behavior:
\be
{\cal S}_L^{op}(t) \simeq \frac{c_{eff}}{3} \log t +{\rm cst}\;,
\label{S_t_scal1}
\ee
which is a generalization of Eq.(\ref{S_t}). 
%
\begin{figure}[h]
\begin{center}
\includegraphics[width=3.3in,angle=0]{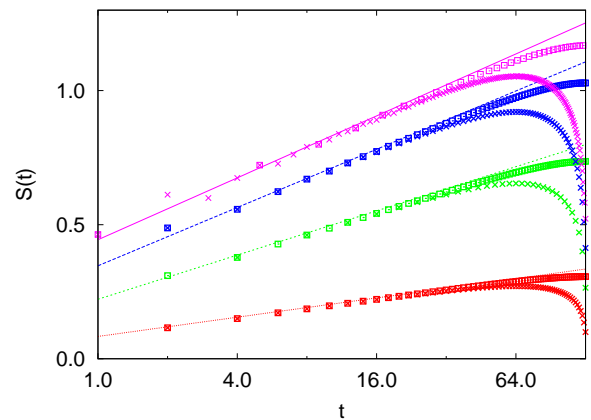}
\end{center}
\caption{
\label{fig:S_logt_loc} (Color online)
Time evolution of the entropy after a quench from two disconnected chains into an
open chain with one defect having different strengths $\Delta=1.$ (homogeneous), $\Delta=.75$,
$\Delta=.5$ and $\Delta=.25$, from the top to the buttom. In the $\log t$ scale the initial
part of the curves for $L=128$ and $L=256$ are close to the indicated straight lines having the slope $c_{eff}(\Delta)/3$, which is calculated
from the scaling of the equilibrium entropy (see Fig.\ref{fig:loc}).}
\end{figure}
%
This relation is checked in Fig.\ref{fig:S_logt_loc} in which the
time evolution of the entropy is shown as a function of $\log t$ for different values of
$\Delta$. Indeed the starting part of the curves are well described by straight lines the
slope of which is compatible with $c_{eff}(\Delta)/3$, as calculated from the equilibrium
entropy and given in Fig.\ref{fig:loc}.

We have repeated the calculation by considering another type of quench: for $t<0$ the system
contains a pair of defects of strength $0<\Delta<1$, which is changed to the homogeneous
coupling, i.e. $\Delta=1$ for $t>0$. In this case the scaling form in Eq.(\ref{S_t_scal})
is still applicable, however with a different effective central charge, $c'_{eff}(\Delta)$,
which is a decreasing function of $\Delta$. We have checked that
$c_{eff}(\Delta)+c'_{eff}(\Delta)=\tilde{c}(\Delta)>1/2$, for example $\tilde{c}(.25)=0.527(3)$,
$\tilde{c}(.5)=0.545(5)$ and $\tilde{c}(.75)=0.520(5)$.

\section{Chain with extended defects}
\label{sec:extended}
\begin{figure}[h]
\begin{center}
\includegraphics[width=3.3in,angle=0]{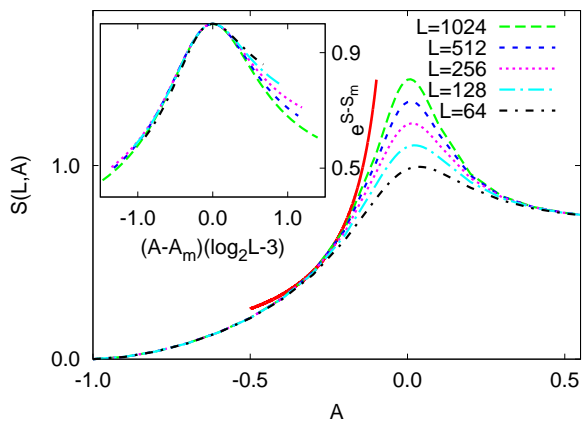}
\end{center}
\caption{
\label{fig:ext} (Color online)
The entanglement entropy of the quantum Ising chain with two symmetrically placed extended
defects as a function of the strength of the defect, $A$, for different finite systems. The
conjectured asymptotic behavior in Eq.(\ref{1perA}) is indicated by the
full red curve. Inset: Scaling plot of the entanglement entropy around its maxima ($\exp(S-S_m)$) in
terms of the combination $(A-A_m)(\log_2 L-3)$, see the text.}
\end{figure}

We have calculated the entanglement entropy between two halves of the quantum Ising chain
which contains a pair of extended defect, as described by the Hamiltonian $H_0+H_{ext}$ in
Eqs.(\ref{eq:H0}) and (\ref{eq:Hext}). For different lengths of the chain, $L$, the
entropy is plotted in Fig.\ref{fig:ext} as a function of the strength of the defect, $A$.

%
%

For a given $L$ the entropy has a maximum close to $A=0$, which corresponds to the critical pure
system, whereas for large negative (positive) $A$ the entropy approaches the limiting
value $0$ ($\log 2$), which is the same in the fully disordered (ordered) phase of the pure system.
For intermediate values of $A \ne 0$ the entropy with increasing $L$ seems to be saturated. This
follows from the observation, that the effective central charges obtained
through Eq.(\ref{two_fit}) approach zero even for a small value of $|A|$.

The finite-size dependence of the entropy can be explained if we take into account the critical scaling
behavior at the interface, which is valid for large enough lengths, $L>\tilde{L}$. As described in Sec.\ref{sec:model}, for $A<0$ the defect renormalizes
to a cut, whereas for $A>0$ it becomes ordered, which is in agreement with the
behavior of the entropy in Fig.\ref{fig:ext}. The microscopic length-scale, $\tilde{L}$, can
be estimated in the $A<0$ regime from the value of the renormalized connecting coupling:
$\left. J_0(L)\right|_{L=\tilde{L}} \sim \tilde{L}^{-|A|}$, which should be in the order of $O(1)$,
say $e^{-\gamma}$, with $\gamma>0$. From this we obtain for the microscopic length:
\be
 \tilde{L} \sim \exp\left(\frac{\gamma}{|A|}\right)\;.
\label{xi_e}
\ee
A similar expression can be derived in the regime $A>0$, too. Note that $\tilde{L}$ has a very fast increase
with decreasing $|A|$ and it is divergent in the homogeneous system. The microscopic length, $\tilde{L}$,
sets in a length scale for the entanglement, too, and in the limit $\tilde{L} \ll L$ the entanglement
entropy is obtained from Eq.(\ref{S_xi}) by replacing $\xi$ with $\tilde{L}$, so that
\be
{\cal S}(A) \simeq \frac{c}{3} \log \tilde{L}+{\rm cst}\simeq \frac{c}{3} \frac{\gamma}{|A|}+{\rm cst}\;.
\label{1perA}
\ee
We have tried to fit the extrapolated numerical values of ${\cal S}(A)$
by this formula in Fig.\ref{fig:ext}, which is found to be reasonable with $\gamma \simeq 0.87$.

We have also analyzed the finite-size scaling behavior of the entropy close to its maximum,
which is located at $A_m=A_m(L)$ and has a value ${\cal S}_m={\cal S}_m(L)$. In order to shift
the maximum to the same position for all $L$ we have considered the difference,
$\Delta {\cal S}(L) \equiv {\cal S}(L)-{\cal S}_m$ as a function of $\Delta A \equiv A-A_m$.
In terms of the scaling variable $\Delta A(\log L -3)$ the curves for different $L$ are
scaled together, as illustrated in the inset of Fig.\ref{fig:ext}. Here the scaling collapse is very good
for $A<0$, whereas for $A>0$ the somewhat less perfect collapse is probably due to the presence
of interface order in the system. Since $A_m(L) \to 0$ for large $L$ the scaling combination
used in the inset of Fig.\ref{fig:ext} is compatible with the microscopic length-scale defined in Eq.(\ref{xi_e}).

\subsection{Time evolution after the quench}

%
\begin{figure}[h]
\begin{center}
\includegraphics[width=3.3in,angle=0]{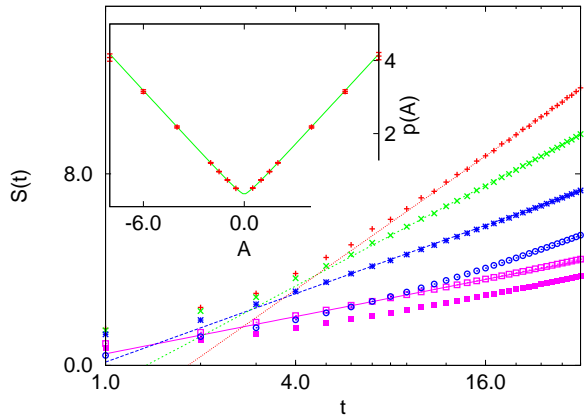}
\end{center}
\caption{
\label{fig:hvl} (Color online)
Time evolution of the entropy after removing a pair of extended defect of various strength, $A$,
from the critical quantum Ising chain (for $t>8.$ from the top to the bottom A=-8.;-6.;-4.;4.;-2.;2.).
In a $\log t$ plot the initial part of the curves
are described by straight lines with $A$ dependent slopes. The extrapolated values of the
slopes as a function of $A$ are given in the inset together with an
interpolation curve, see the text.}
\end{figure}
%
We have measured the time evolution of the entropy, if the system contains a pair of
extended defects for $t<0$, which is removed for $t>0$ and we are left with the homogeneous closed chain.
As shown in Fig.\ref{fig:hvl} the entropy for $t \ll L$ has a logarithmic increase in time and the prefactor, $p(A)$, which
seems to have the symmetry: $p(A)=p(-A)$ is increasing with $|A|$. We have calculated
two-point fits for the prefactors, see in Eq.(\ref{two_fit_t}),
the extrapolated values of which are plotted in the inset of Fig.\ref{fig:hvl}. 
The prefactor has its minimum around $A=0$, which is close to $2c/3=1/3$, whereas for large
$|A|$ we have approximately: $p(A) \approx A/2$. The measured points can be well interpolated
by the curve: $p(A)=[1/9+|A|/6+A^2/4]^{1/2}$, which is also indicated in the inset of Fig.\ref{fig:hvl}.

To explain the observed behavior of the entropy we have to take into account two different effects
of the extended defect. First, from local scaling consideration in Sec.\ref{sec:model} the defect
renormalizes to a cut, thus - for small $|A|$ - the quench is made from two disconnected parts
to a closed chain, and according to Eq.(\ref{S_t_conf}) the prefactor is $2c/3=1/3$ in agreement
with the measured limiting values. The second effect of the extended defect is to produce
an increase of the entropy through inhomogeneous quench, since the interaction in the chain
at position $x$ differs from the critical value by $J(x)-1$. According to the argument in Ref.\cite{CC05}
pairs of quasiparticles are emitted at different points of the chain and they will contribute
to the entanglement at later time, when reach the two parts of the system.
The density of quasiparticles, $\alpha(x)$ depends on the distance from the critical point at
the given position\cite{Fagotti08}, and for a small perturbation it is given by\cite{eip09} $\alpha \simeq |J(x)-1|/2$.
The increase of the entropy is obtained by integrating over the contributions:
\be
{\cal S}(t)-{\cal S}(0)=\frac{1}{2} \int_{-t}^{t} {\rm d} x \alpha[J(x)]\;,
\label{S_t_int}
\ee
which for the smooth inhomogeneity in Eq.(\ref{eq:Hext}) and for $t \ll L$ results in:
${\cal S}(t)-{\cal S}(0) \approx A/2 \log t$, in agreement with the numerical results.

\section{Discussion}

\label{sec:disc}

In this paper we have considered the quantum Ising chain with critical couplings which
is separated into two parts by localized or extended defects. In both cases the scaling behavior
of the interface magnetization is non-universal (the scaling exponents depend on the
strength of the defect) and we asked the question, how this fact is reflected in the
entanglement properties of the system. For the localized defect both the equilibrium
and the nonequilibrium entropy is found to be characterized by the same effective
central charge, the value of which depends on the strength of the defect. In this case scaling
of the equilibrium and the nonequilibrium entropy can be cast into the same form, see
Eq.(\ref{S_t_scal}).

The situation is found completely different for the extended defect, in which case the
equilibrium entropy is saturated, although the magnetization correlations (also at the
interface) are long ranged and thus the corresponding correlation length is divergent. In
this case the entanglement is related to another, finite microscopic length, as given in
Eq.(\ref{xi_e}). This is due to the fact that the two parts of the system at the interface are
asymptotically separated and the wave-function become localized. We can thus
conclude that for an extended defect the correlation length and the entanglement length
have different scaling properties. As far as the nonequilibrium entropy of this system
is concerned it is shown to have a logarithmic $t$-dependence, the prefactor of which
is the result of two effects: the asymptotic cut and the inhomogeneous quench.

Our results for the localized defect are related to similar studies for the $XX$-chain\cite{Peschel05,Levine04}.
Since the entropy of the quantum Ising chain and that of the $XX$-chain are exactly
related\cite{IJ07} the same is true for the central charges, too. For example from $c_{eff}(\Delta)$
in Fig.\ref {fig:loc} we obtain the effective central charge in the closed $XX$-chain with
one interface defect of strength $t=\Delta$ as $c^{XX}_{eff}(t)=c_{eff}(\Delta)+1/2$, as given
in Fig.5 of Ref.\cite{Peschel05}. Here we comment on the observation
in Ref.\cite{Peschel05} that the small $\Delta$ dependence of the effective central charge is given by:
$c_{eff}(\Delta) \sim \Delta^{\delta}$, with an exponent, $\delta \approx 1.8$. According
to our perturbative calculation in Eq.(\ref{c_eff_Delta}) the true exponent is $\delta=2$,
however with a multiplicative logarithmic correction term\cite{note}. Similar behavior is expected to
hold in higher dimensional gapless fermionic systems with weak links\cite{Levine08}.
The result in Eq.(\ref{c_eff_Delta})
can be compared with a bosonization study of the continuum version of the XX-chain\cite{Levine04},
in which the impurity (defect) contribution to the entanglement entropy is found to scale as:
\be
\delta {\cal S}=\frac{1}{4} y^2 \epsilon^2 \ln \frac{L}{\epsilon}\left(1-\ln \frac{L}{\epsilon}\right)\;,
\label{imp}
\ee
where $y$ is the strength of the impurity potential and $\epsilon$ is a small-distance cutoff.
The leading $\ln^2(L/\epsilon)$ term in Eq.(\ref{imp}) is not
compatible with our lattice result.

As far as marginal extended defects are concerned our results for the nonequilibrium entropy
are expected to be generic for another critical quantum spin chains, too. From scaling
theory it is known that an extended defect in the form, $(A/x)^{\omega}$, $x$ being
the distance from the center of the defect, is a marginal perturbation, provided $\omega=1/\nu$,
$\nu$ being the correlation length exponent\cite{ipt}. Now estimating the nonequilibrium entropy
we repeat the argument at the end of Sec.\ref{sec:extended}, where the density of
quasiparticles is expected to scale with the value of the local gap\cite{Sotiriadis08}:
$\alpha(x) \sim
|J(x)-1|^{\nu} \sim A/x$. Integrating the contributions in time, see Eq.(\ref{S_t_int}),
for large $A$ leads to the behavior, ${\cal S}(t)-{\cal S}(0) \sim A \log t$,
as for the quantum Ising chain.

This work has been
 supported by the Hungarian National Research Fund under grant No OTKA
TO48721, K62588, K75324 and MO45596 and by a German-Hungarian exchange program (DFG-MTA).
We are grateful to I. Peschel and H. Rieger for useful discussions.

\appendix

\section{Time evolution of the entropy for quadratic fermionic systems}
\label{sec:quadr}

Let us consider a general Hamiltonian, $\cal H$, which is
quadratic in terms of fermion creation, $c_k^{\dag}$, and annihilation, $c_k$,
operators and which is given for $t \ge 0$ as:
\be
{\cal H}=\sum_{k,l=1}^L\left[ c_k^{\dag}{  A}_{kl}c_l +\frac{1}{2} \left(
c_k^{\dag}{  B}_{kl}c_l^{\dag} + {\rm h.c.} \right)\right] \;.
\label{Hquad}
\ee
Here ${  A}_{kl} \equiv (\mathbf{A})_{kl}={  A}_{lk}$ and ${  B}_{kl}\equiv (\mathbf{B})_{kl}=-{  B}_{lk}$ are real numbers and
$k,l$ are the sites of a lattice. In the initial state, i.e. for $t<0$ the parameters of the
Hamiltonian are different, say, ${A}_{kl}^{(0)}$ and ${B}_{kl}^{(0)}$ and the ground state of
the initial Hamiltonian, ${\cal H}^{(0)}$,
is denoted by $|\psi_0 \rangle$. The system is divided into two parts: ${\cal A}$ consists of
points $k=1,2,\dots,\ell$ and ${\cal B}$ of the rest of the system.

For one dimensional spin models, such as the quantum Ising chain, the Pauli
spin-operators, $\sigma_l^{x,y,z}$, are related to the fermionic operators as\cite{lsm}:
\beqn
\prod_{j<l}(-\sigma_j)^z \sigma_l^x=c_l^{\dag}+c_l\equiv {\mathfrak A}_l \equiv (-1)^{l-1} \check{a}_{2l-1}
 \cr
\prod_{j<l}(-\sigma_j)^z \imath \sigma_l^y=c_l^{\dag}-c_l\equiv {\mathfrak B}_l \equiv (-1)^{l-1} \check{a}_{2l}\;.
\label{AB}
\eeqn
Here, in the last two equations, at a given site, $l$, we define two Clifford operators, ${\mathfrak A}_l$
and ${\mathfrak B}_l$, as well as two Majorana fermion operators, $\check{a}_{2l-1}$ and
$\check{a}_{2l}$. The commutation relations for these new set of operators are:
\beqn
{\mathfrak A}_l^2=1,\quad {\mathfrak B}_l^2=-1,\quad {\mathfrak A}_l{\mathfrak B}_k=-{\mathfrak B}_k{\mathfrak A}_l, \cr
{\mathfrak A}_l{\mathfrak A}_k=-{\mathfrak A}_k{\mathfrak A}_l,\quad {\mathfrak B}_l{\mathfrak B}_k=-{\mathfrak B}_k{\mathfrak B}_l,\quad
l \neq k
\label{comm_cl}
\eeqn
for the Clifford operators and
\be
\check{a}_{l}^{+}=\check{a}_{l},\quad \{ \check{a}_{l},\check{a}_{k} \}=2\delta_{l,k}
\label{comm_maj}
\ee
for the Majorana fermion operators.

The {\it first step} of the calculation is to diagonalize both , ${\cal H}^{(0)}$ and ${\cal H}$,
which can be made by the same type of canonical transformation. For simplicity we work here with ${\cal H}$,
for ${\cal H}^{(0)}$ the analogous results are denoted by a superscript $^{(0)}$.
The new set of fermion operators are given by the combination:
\be
\eta_k=\sum_l(g_{kl} c_l + h_{kl} c_l^{\dag})
\ee
where $g_{kl}$ and $h_{kl}$ are real numbers and
the Hamiltonian assumes the diagonal form:
\be
{\cal H}=\sum_{k=1}^L \Lambda_k \eta_k^{\dag} \eta_k + {\rm const.}\;,
\label{H_free}
\ee
Here the energy of the free-fermionic modes, $\Lambda_k$, are given
by the solution of the eigenvalue equations:
\beqn
(\mathbf{A}-\mathbf{B})(\mathbf{A}+\mathbf{B})\mathbf{\Phi}_k&=&\Lambda_k^2 \mathbf{\Phi}_k \cr
(\mathbf{A}+\mathbf{B})(\mathbf{A}-\mathbf{B})\mathbf{\Psi}_k&=&\Lambda_k^2 \mathbf{\Psi}_k\;.
\label{Lambda}
\eeqn
and the components of the eigenvectors are:
\beqn
\Phi_k(i)=g_{ki}+h_{ki} \cr
\Psi_k(i)=g_{ki}-h_{ki}\;.
\eeqn
In the {\it second step} we consider the subsystem ${\cal A}$ and calculate its reduced density
matrix, $\rho_A$, which can be reconstructed from the time-dependent reduced correlation matrix
of the Majorana operators:
\be
\langle \psi_0|\check{a}_{m}(t)\check{a}_{n}(t)|\psi_0 \rangle = \delta_{m,n} + \imath (\Gamma^A_{\ell})_{mn}\;,
\label{corr_maj}
\ee
$m,n=1,2,\dots,2\ell$. Here $(\Gamma^A_{\ell})_{mn}$ is a skew-symmetric (or antisymmetric) matrix which
is transformed by an orthogonal transformation, $Q$, into a block-diagonal form:
\be
Q\Gamma^A_{\ell}Q^T=
\begin{bmatrix}
0      &\nu_1& 0    &0     & \dots       &     &\cr
-\nu_1 &0    & 0    &0     & \dots       &     &\cr
0      &0    & 0    &\nu_2 & \dots       &     &\cr
0      &0    &-\nu_2&0     & \dots       &     &\cr
       &     &      &      & \ddots      &     &\cr
       &     &      &      & \dots0      &\nu_r&\cr
       &     &      &      & \dots-\nu_r &0    &\cr
       &     &      &      &             &     &\ddots
\end{bmatrix}
\ee
thus the eigenvalues of $\Gamma^A_{\ell}$ are $\pm \imath \nu_r$, $r=1,2,\dots,\ell$.
In this representation the reduced density matrix is the direct
product of $\ell$ uncorrelated modes: $\rho_\ell=\bigotimes_{r=1}^{\ell} \rho_r$,
where $\rho_r$ has eigenvalues
$(1 \pm \nu_r)/2$. The entanglement
entropy is then given by the sum of binary entropies:
\be
S_L(\ell)=-\sum_{r=1}^{\ell} \left(\frac{1+\nu_r}{2} \log_2 \frac{1+\nu_r}{2}
+\frac{1-\nu_r}{2} \log_2 \frac{1-\nu_r}{2}\right).
\label{binary}
\ee
In the {\it third step} we calculate the time-dependent correlation matrix and work in terms
of the Clifford-operators the time evolution of which are given by\cite{IgloiRieger00}:
\beqn
{\mathfrak A}_l(t)=\sum_k\left[ \langle A_l A_k \rangle_t {\mathfrak A}_k +
 \langle A_l B_k \rangle_t {\mathfrak B}_k\right] \cr
{\mathfrak B}_l(t)=\sum_k\left[ \langle B_l A_k \rangle_t {\mathfrak A}_k +
 \langle B_l B_k \rangle_t {\mathfrak B}_k\right]
\eeqn
Here the time-dependent contractions are:
\beqn
\langle A_l A_k \rangle_t&=&\sum_q \cos( \Lambda_q t) \Phi_q(l) \Phi_q(k)\; ,\nonumber\\
\langle A_l B_k \rangle_t&=&\langle B_k A_l \rangle_t=i \sum_q \sin (\Lambda_q t) \Phi_q(l) \Psi_q(k)\; ,\nonumber\\
\langle B_l B_k \rangle_t&=&\sum_q \cos (\Lambda_q t) \Psi_q(l) \Psi_q(k)\; .
\label{contr}
\eeqn
Note, that in Eq.(\ref{contr}) the free-fermionic quantities are related to the Hamiltonian $\cal H$, which
governs the time evolution in the system for $t>0$. The matrix-elements of the time-dependent Clifford
operators, such as $\langle \psi_0|{\mathfrak A}_l(t){\mathfrak A}_k(t)|\psi_0 \rangle$, involve the following ground-state expectation values:
\beqn
\langle \psi_0|{\mathfrak A}_{k_1}{\mathfrak A}_{k_2}|\psi_0 \rangle=\delta_{k_1,k_2},\ 
\langle \psi_0|{\mathfrak B}_{k_1}{\mathfrak B}_{k_2}|\psi_0 \rangle=-\delta_{k_1,k_2} \cr
\langle \psi_0|{\mathfrak A}_{k_1}{\mathfrak B}_{k_2}|\psi_0 \rangle=-G^{(0)}_{k_2k_1},\
\langle \psi_0|{\mathfrak B}_{k_1}{\mathfrak A}_{k_2}|\psi_0 \rangle=G^{(0)}_{k_1k_2}\;.
\eeqn
Here the first equations follow from the commutation rules in Eq.(\ref{comm_cl}), whereas the
static correlation matrix $G^{(0)}_{k_1k_2}$ is given by:
\be
G^{(0)}_{k_1k_2}=-\sum_q \Psi_q^{(0)}(k_1) \Phi_q^{(0)}(k_2)\;,
\ee
which is calculated with the initial Hamiltonian, ${\cal H}^{(0)}$. Now one can go on
and calculate the time-dependent
correlation matrix of the Clifford operators, which is then transformed by Eq(\ref{AB}) for Majorana operators
so that finally one obtains the matrix-elements of $\Gamma^A$ in Eq.(\ref{corr_maj}) as:
\begin{widetext}
\beqn
\Gamma^A_{2l-1,2m-1}=-\imath \left[\sum_{k_1,k_2} G^{(0)}_{k_1k_2} \langle A_l B_{k_1} \rangle_t
\langle A_m A_{k_2} \rangle_t
-\sum_{k_1,k_2} G^{(0)}_{k_2k_1} \langle A_l A_{k_1}\rangle_t \langle A_m B_{k_2} \rangle_t \right](-1)^{l+m} \cr
\Gamma^A_{2l-1,2m}= \left[\sum_{k_1,k_2} G^{(0)}_{k_2k_1} \langle A_l A_{k_1} \rangle_t
\langle B_m B_{k_2} \rangle_t
-\sum_{k_1,k_2} G^{(0)}_{k_1k_2} \langle A_l B_{k_1} \rangle_t \langle B_m A_{k_2}\rangle_t \right](-1)^{l+m} \cr
\Gamma^A_{2l,2m-1}= -\left[\sum_{k_1,k_2} G^{(0)}_{k_2k_1} \langle A_m A_{k_1} \rangle_t
\langle B_l B_{k_2} \rangle_t
-\sum_{k_1,k_2} G^{(0)}_{k_1k_2} \langle A_m B_{k_1} \rangle_t \langle B_l A_{k_2}\rangle_t \right](-1)^{l+m} \cr
\Gamma^A_{2l,2m}=-\imath \left[\sum_{k_1,k_2} G^{(0)}_{k_2k_1} \langle B_l A_{k_1} \rangle_t
\langle B_m B_{k_2} \rangle_t
-\sum_{k_1,k_2} G^{(0)}_{k_1k_2} \langle B_l B_{k_1}\rangle_t \langle B_m A_{k_2} \rangle_t \right](-1)^{l+m}
\eeqn
\end{widetext}

\section{Perturbative calculation of the entropy for localized defects}
\label{sec:pert}
Here we consider the homogeneous quantum Ising chain with one single defect coupling of strength, $\Delta$,
between spins at $\ell$ and $\ell+1$, but the two ends of the chain at $1$ and $L$ are free. The unperturbed
system (with $\Delta=0$) consists of two separated chains: ${\cal A}$ with sites $i=1,2,\dots \ell$ and
${\cal B}$ with sites $i=\ell+1,\ell+2,\dots L$, and the length of ${\cal B}$ is denoted by $\ell'=L-\ell$.
The reduced density matrix of ${\cal A}$, denoted by ${\mathbf \rho}_{\ell}$, is calculated perturbatively
in Ref.\cite{IJ07}. The matrix-elements in leading order, i.e. up to $O(\Delta^2)$ are given by:
\beqn
\langle \varphi_{i}^{\cal A}|{\mathbf \rho}_{\ell}|\varphi_{j}^{\cal A}\rangle={\mathbf \rho}_{\ell}(i,j)
={\cal Z}\sum_k^{\cal B} c(i,k)c^*(j,k)\;,
\label{rho}
\eeqn
with $c(0,0)=1$ and
\be
c(i,k)=\frac{\Delta}{E_i^{\cal A}+E_k^{\cal B}}\langle\varphi_{i}^{\cal A}|\sigma_{\ell}^x|\varphi_{0}^{\cal A}\rangle
 \langle\varphi_{k}^{\cal B}|\sigma_{\ell+1}^x|\varphi_{0}^{\cal B}\rangle\;.
\label{factor}
\ee
Here $|\varphi_{i}^{\cal A}\rangle$ ($|\varphi_{k}^{\cal B}\rangle $) is the $i$-the ($k$-th) eigenstate
of the unperturbed system ${\cal A}$ (${\cal B}$) with excitation energy: $E_i^{\cal A}$ ($E_k^{\cal B}$)
and ${\cal Z}$ is a normalization constant, so that $\sum_i^{\cal A}{\mathbf \rho}_{\ell}(i,i)=1$.
In Eq.(\ref{factor}) the matrix-elements of
the surface magnetization operators $\sigma_{\ell}^x$ of system ${\cal A}$ and $\sigma_{\ell+1}^x$
of system ${\cal B}$ are non-zero, if the excited states, $|\varphi_{i}^{\cal A}\rangle$ and
$|\varphi_{k}^{\cal B}\rangle$ in the fermionic representation contain just one fermion. Then using
the notation of App.\ref{sec:quadr} the
non-vanishing matrix-elements are given by: $\langle\varphi_{i}^{\cal A}|\sigma_{\ell}^x|\varphi_{0}^{\cal A}\rangle=\Psi_i^{\cal A}(\ell)$, $i=1,2,\dots \ell$ with $E_i^{\cal A}=\Lambda_i^{\cal A}$
and $\langle\varphi_{k}^{\cal B}|\sigma_{\ell+1}^x|\varphi_{0}^{\cal B}\rangle=\Psi_k^{\cal B}(1)$,
$k=1,2,\dots \ell'$ with $E_k^{\cal B}=\Lambda_k^{\cal B}$. Using the exact solution of the open
quantum Ising chain with homogeneous critical couplings of length $\ell$ (see Eqs.(A.5) and (A.6) of
Ref.\cite{IgloiLin08}) we obtain in leading order:
\beqn
{\mathbf \rho}_{\ell}(i,j)&=&{\cal Z}\Delta^2
\frac{\cos\alpha_i\cos\alpha_j} {(2\ell+1)(2\ell'+1)}\cr
&\times& \sum_{k=1}^{\ell'}\frac{\cos^2\beta_k}{(\sin \alpha_i + \sin \beta_k)(\sin \alpha_j + \sin \beta_k)}
\eeqn
$i,j=1,2,\dots,\ell$, with $\alpha_i=\frac{\pi}{2}\frac{2i-1}{2\ell+1}$
and $\beta_k=\frac{\pi}{2}\frac{2k-1}{2\ell'+1}$. In the following we analyse the consequences of this
expression in the limits: $\ell \gg 1$ and $\ell'/\ell \gg 1$.

The leading eigenvalue of the reduced density matrix is:
\be
w_1={\cal Z}=1-\Delta^2\left(a_1 + a \log \ell \right)\;,
\label{w0}
\ee
where the prefactor of the logarithm is $a=\pi^{-2}$, and $a_1$ is a constant of $O(1)$.
From Eq.(\ref{w0}) we obtain for the single-copy entanglement:
\be
{\cal S}_1=\Delta^2 \left( \frac{1}{\pi^2} \log \ell + {\rm cst.}\right)
\ee
which is proportional to $\log \ell$. For {\it two} symmetrically placed
defects, as studied numerically in Sec.\ref{sec:localised} the prefactor is $2\Delta^2/\pi^2$, from
which we obtain the value of $\kappa_{eff}(\Delta)$ as given in Eq.(\ref{kappa_eff}).

To obtain the entanglement entropy in leading order one should calculate the other $i=2,3,\dots \ell+1$
eigenvalues of the reduced density matrix, which are expressed as: $w_i=\Delta^2 \epsilon_i$, so that
the entropy is given by:
\be
{\cal S}(\Delta)=-w_1 \log w_1 -\Delta^2\sum_{i=2}^{\ell+1} \epsilon_i(\log \epsilon_i
+ \log \Delta^2)+O(\Delta^4)\;.
\label{S_Delta}
\ee
Here the correction term is evaluated numerically and we have obtained
$\sum_{i=2}^{\ell+1} \epsilon_i \log \epsilon_i=b_1 + b \log \ell$
and the prefactor of the logarithm is $b=0.062180(2)$. Putting this and $w_1$ from Eq.(\ref{w0})
into Eq.(\ref{S_Delta}) we obtain that the entanglement entropy scales as $\log \ell$ for
large $\ell$, and the effective central charge (calculated for {\it two} symmetrically placed
defects) is given in Eq.(\ref{c_eff_Delta}).


\end{document}